\documentclass[prl,aps,graphics,floats,showpacs,preprint]{revtex4}
\usepackage{graphicx}
\usepackage{amssymb}
\usepackage{amsmath}
\begin{document}

\draft

\title{Non-monotonic variation of anomalous Hall conductivity with spin orbit coupling strength}

\author{M. Chen, Z. Shi, and S. M. Zhou}
\address{Surface Physics Laboratory (State Key Laboratory) and Department of Physics, Fudan University, Shanghai 200433, China}
\author{J. Li}
\address{Department of Optical Science and Engineering, Fudan University, Shanghai 200433, China}
\author{W. J. Xu}
\address{Department of Physics, Hong Kong University of Science and Technology, Kowloon, Hong Kong, China}
\author{X. X. Zhang}
\address{Image-characterization Core Lab, 4700 King Abdullah University of Science and Technology (KAUST), Thuwal 23900-6900, Kingdom of Saudi Arabia}
\author{J. Du}
\address{National Laboratory of Solid State Microstructures, Nanjing University, Nanjing 210093, China}

\date{\today}

\begin{abstract}

For L1(0) FePt films, the anomalous Hall resistivity is found to
be proportional to spontaneous magnetization $M_{\mathrm{S}}$.
After the $M_{\mathrm{S}}$ temperature effect is eliminated,
$\rho_{\mathrm{xyo}}$ can be fitted by
$\rho_{\mathrm{xyo}}=a_{\mathrm{o}}\rho_{\mathrm{xx}}+
b_{\mathrm{o}}\rho_{\mathrm{xx}}^{\mathrm{2}}$. $a_{\mathrm{o}}$
and $b_{\mathrm{o}}$ change non-monotonically with  chemical long
range ordering degree $S$. Accordingly, it is indicated that for
L1(0) FePt films the spin orbit coupling strength increases
monotonically with increasing $S$.

\end{abstract}

\vspace{0.5 cm} 
\maketitle

\indent Anomalous Hall effect (AHE) in ferromagnetic metallic
films has been studied extensively because of its intriguing
physics~\cite{1,2,3,4,5,6,7,8,9,10,11,12,13,14,15,29}. It is shown
both theoretically and experimentally that the AHE resistivity is
fitted by $\rho_{\mathrm{xy}}=a\rho_{\mathrm{xx}}+ b\rho_{\mathrm{
xx}}^{\mathrm{2}}$ with the longitudinal resistivity
$\rho_{\mathrm{xx}}$. One has the anomalous Hall conductivity
(AHC) $\sigma_{\mathrm{xy}}\simeq-a\sigma_{\mathrm{xx}}-b$ when
$\rho_{\mathrm{xy}}\ll\rho_{\mathrm{xx}}$, where
$a=-\sigma_{\mathrm{sk}}/\sigma_{\mathrm{xx}}$ with
$\sigma_{\mathrm{xx}}=1/\rho_{\mathrm{xx}}$,
$b=-\sigma_{\mathrm{sj}}-\sigma_{\mathrm{int}}$,
$\sigma_{\mathrm{sk}}$ arises from the extrinsic skew scattering
at impurity sites, $\sigma_{\mathrm{sj}}$ and
$\sigma_{\mathrm{int}}$ correspond to the extrinsic side-jump
scattering and the intrinsic Karplus-Luttinger terms,
respectively~\cite{2,3,4}. Very recently, $\sigma_{\mathrm{int}}$
has been recalculated by the integration of the Berry curvature
$\Omega(\vec{k})$ of the Bloch states in the Brillouin
zone~\cite{5,6},
\begin{equation}
\sigma_{\mathrm{int}}=-\frac{e^2}{\hbar}\int\frac{d^3k}{(2\pi)^3}{\Omega}^z(\vec{k})
\label{paul1}
\end{equation}
\indent It is theoretically shown that as caused by the spin orbit
coupling (SOC), $\sigma_{\mathrm{sk}}$ and $\sigma_{\mathrm{sj}}$
exhibit non-monotonic variations with the SOC strength
$\Delta_{\mathrm{SOC}}(=\xi\vec{l}\cdot\vec{s})$ whereas
$\sigma_{\mathrm{int}}$ demonstrates monotonic
variation~\cite{6,14,15}. There is still a lack of experimental
evidence because most such experimental studies are focused on
3\emph{d} transition metallic films~\cite{7,8,9,10,11} and
$\Delta_{\mathrm{SOC}}$ can be tuned only in a small regime. In
contrast, L1(0) FePt films, as important magnetic recording media,
may be an ideal object for this purpose because
$\Delta_{\mathrm{SOC}}$ can be tuned by the chemical long range
ordering degree $S$, as demonstrated by the enhancement of
magneto-optical Kerr effect (MOKE) and spin Hall
effect~\cite{16,19}. More importantly, up to date, the SOC
mechanism remains unclear although it is of crucial importance for
the magnetism of L1(0) FePt films. For example,
$\Delta_{\mathrm{SOC}}$ in L1(0) FePt films strongly depends on
the orbital polarization of Pt atoms but experimental results
about the magnitude of the orbital polarization are
controversial~\cite{39,40}. Studies of the AHE in L1(0) FePt films
as a function of $S$ are helpful to deeply understand the nature
of both the AHE and in particular the SOC in L1(0) FePt films. \\
\indent In this work, we have studied the SOC effect on the AHE by
employing L1(0) FePt films, in which the phase transformation and
thus $S$ are easily controlled by varying either deposition or
post-annealing conditions. $\rho_{\mathrm{xy}}$ is found to change
in a linear scale of spontaneous magnetization $M_{\mathrm{S}}$
below 300 K. After the thermally driven $M_{\mathrm{S}}$ reduction
is considered, $\rho_{\mathrm{xyo}}$ can be fitted in a scale of
$\rho_{\mathrm{xyo}}=a_{\mathrm{o}}\rho_{\mathrm{xx}}+b_{\mathrm{o}}\rho_{\mathrm{xx}}^{\mathrm{2}}$.
It is interesting to find that $a_{\mathrm{o}}$ and
$b_{\mathrm{o}}$\emph{ both} change non-monotonically with $S$.
Based on two dimensional electron gas model~\cite{14,15} and other
calculations~\cite{6}, the
present AHE experiments indicate the SOC enhancement in  L1(0) FePt films.  \\
\indent A series of 10 nm thick L1(0)
Fe$_{\mathrm{50}}$Pt$_{\mathrm{50}}$ (=FePt) films with altering
$S$ were grown on MgO(001) substrates by DC magnetron sputtering
at different substrate temperatures. The microstructure and the
film thickness were identified by x-ray diffraction (XRD) and
reflectometry (XRR), respectively. The films were patterned into
normal Hall bar and the Hall resistance $\rho_{\mathrm{H}}$ was
measured from 5 K to 300 K. The longitudinal resistance
$\rho_{\mathrm{xx}}$ was also measured in the same temperature
regime at zero external magnetic field. In experiments, the
magnetoresistance of all samples is less than 0.5$\%$.
$M_{\mathrm{S}}$ was measured as a function of temperature by
PPMS. Polar MOKE spectra
were measured at room temperature by a home-made Kerr spectrometer~\cite{23}.\\
\indent Deposited at ambient temperature, the FePt film is of
disordered fcc structure with (111) preferred orientation. At high
substrate temperatures, peaks begin to appear near 24 degrees and
48 degrees corresponding to L1(0) phase (001) and (002)
orientations, as shown in Fig.~\ref{Fig1}(a). It is found that
when the substrate temperature is increased, $S$ as calculated
from the intensities of (001) and (002) peaks~\cite{24} increases
from 0 to 0.86 and the lattice constant along the $c$ axis
decreases from 0.382 nm to 0.374 nm. Therefore, the long range
chemical ordering and the lattice distortion happened
simultaneously. Due to its crucial importance for calculations of
$\rho_{\mathrm{xy}}$ and $\rho_{\mathrm{xx}}$, the film thickness
was measured by XRR at low angles, as shown in
Fig.~\ref{Fig1}(b), and found to be $10 \pm 0.5$ nm.\\
\indent Figure~\ref{Fig1}(c) shows typical Hall loops at 5 K. For
$S=0$, the Hall loop is slanted with hard axis along the film
normal direction. For large $S$, the loop becomes squared with
large coercivity. Apparently, the perpendicular magnetic
anisotropy is established in L1(0) FePt films. For the Hall loop
of ferromagnetic films, Hall resistivity
$\rho_{\mathrm{H}}=R_{\mathrm{O}}H + 4\pi M(H)R_{\mathrm{S}}$,
where $R_{\mathrm{O}}$ and $R_{\mathrm{S}}$ are coefficients of
ordinary and anomalous Hall effects, respectively. By
extrapolating the saturation curve of $\rho_{\mathrm{H}}$ versus
$H$, the AHE resistivity $\rho_{\mathrm{xy}}$ is achieved and
found to decrease for large $S$~\cite{25}. As shown in
Fig.~\ref{Fig1}(d), for $S=0$ and 0.86 the normalized spontaneous
magnetization decreases by about $15 \%$ with increasing
temperature from 5 K to 300 K. Apparently, the $M_{\mathrm{S}}$
reduction cannot be ignored because the Curie temperature of
700-750 K is not sufficiently high~\cite{18}. Moreover, for $S=0$
and 0.86, $M_{\mathrm{S}}$ changes in a linear scale of
$T^{\mathrm{2}}$, hinting either the excitation of interacting
spin waves or long-wavelength, low-frequency
fluctuations~\cite{9,30}. For all samples, one has
$M_{\mathrm{S}}=M_{\mathrm{o}}f(T)$, where $M_{\mathrm{o}}$ and
$f(T)$ are the spontaneous magnetization at zero temperature and
the temperature dependent factor, respectively. \\
\indent Figures~\ref{Fig2}(a) and~\ref{Fig2}(b) show that
$\rho_{\mathrm{xx}}$ and $\rho_{\mathrm{xy}}$ both increase with
temperature but decrease with $S$. $\rho_{\mathrm{xx}}$ approaches
the residual resistance $\rho_{\mathrm{o}}$ near zero temperature
and the latter becomes small for high $S$ possibly due to both
improvement of the crystalline quality and reduction of the
density of static defects at elevated substrate temperatures.
Figures~\ref{Fig2}(c) and~\ref{Fig2}(d) show typical curves of
$\rho_{\mathrm{xy}}/\rho_{\mathrm{xx}}$ versus
$\rho_{\mathrm{xx}}$ as a function of temperature. For all
samples, the curves of $\rho_{\mathrm{xy}}/\rho_{\mathrm{xx}}$
versus $\rho_{\mathrm{xx}}$ have downward curvatures and cannot be
fitted with the formula $\rho_{\mathrm{xy}}/\rho_{\mathrm{xx}}
=a+b\rho_{\mathrm{xx}}$. In order to reveal the relationship
between $\rho_{\mathrm{xy}}$ and $\rho_{\mathrm{xx}}$, they are
often measured as a function of temperature. At the same time,
$M_{\mathrm{S}}$ generally decreases with increasing temperature
for ferromagnetic materials with low Curie temperature. If
$\rho_{\mathrm{xy}}$ is proportional to $M_{\mathrm{S}}$, one has
the following equation.
\begin{equation}
\rho_{\mathrm{xy}}=\rho_{\mathrm{xyo}}f(T) \label{paul3}
\end{equation}
As such, the $f(T)$ independent AHE resistivity
$\rho_{\mathrm{xyo}}$ can be fitted by
$\rho_{\mathrm{xyo}}/\rho_{\mathrm{
xx}}=a_{\mathrm{o}}+b_{\mathrm{o}}\rho_{\mathrm{xx}}$. As shown in
Figs.~\ref{Fig2}(c) and~\ref{Fig2}(d),
$\rho_{\mathrm{xyo}}/\rho_{\mathrm{ xx}}$ can be fitted by a
linear function of $\rho_{\mathrm{ xx}}$. Such salient linear
dependence indicates that for FePt films $\rho_{\mathrm{xy}}$ is
proportional to $M_{\mathrm{S}}$ in the sampling temperature
region. Similar phenomena have been observed in Ni alloys,
Heuslers(CoMnSb, NiMnSb, and Co$_{\mathrm{2}}$CrAl), Si-based
magnetic semiconductor, and other ferromagnetic
compounds~\cite{9,31,32,33,34,35,36}. It has been pointed out that
the skew scattering contribution has linear dependence on
$M_{\mathrm{S}}$~\cite{14,28}. As a result, one has
$a=a_{\mathrm{o}}f(T)$ and $b=b_{\mathrm{o}}f(T)$, where
$a_{\mathrm{o}}$ and $b_{\mathrm{o}}$ are $f(T)$ independent. The
intrinsic $\sigma_{\mathrm{int}}$ is proved to have linear
dependence on $M_{\mathrm{S}}$ by both the Karplus-Luttinger model
and the integration of the Berry curvature~\cite{2,9}, so does
$\sigma_{\mathrm{sj}}$. Therefore, one has
$\sigma_{\mathrm{xyo}}\simeq-a_{\mathrm{o}}\sigma_{\mathrm{xx}}-b_{\mathrm{o}}$,
where
$a_{\mathrm{o}}=-\sigma_{\mathrm{sko}}/\sigma_{\mathrm{xx}}$,
$b_{\mathrm{o}}=-\sigma_{\mathrm{sjo}}-\sigma_{\mathrm{into}}$,
and $\sigma_{\mathrm{xyo(sko,~sjo~or~into)}}=\sigma_{\mathrm
{xy(sk,~sj~or~int)}}/f(T)$. In the following, we will become
concerned about corresponding $f(T)$ independent physical quantities.\\
\indent Figures~\ref{Fig3}(a)~and~\ref{Fig3}(b) show that
$a_{\mathrm{o}}$ and $b_{\mathrm{o}}$ change non-monotonically
with $S$. For $S=0$ and $0.86$, $a_{\mathrm{o}}$ approaches zero
and has a minimal value at intermediate $S$. $b_{\mathrm{o}}$ is
equal to 700 ($\Omega$cm)$^{-1}$ for $S=0$ and increases with a
maximal value of 900 ($\Omega$cm)$^{-1}$, and finally approaches a
saturation value of 600 ($\Omega$cm)$^{-1}$ for $S=0.86$. Opposite
signs of $a_{\mathrm{o}}$ and $b_{\mathrm{o}}$ indicate that the
skew scattering has contribution to the AHC in an opposite way to
those of the side-jump and the Karplus-Luttinger terms, as
observed in bcc Fe films~\cite{11}. Very recently, a new approach
has been proposed to fit the data by Tian \emph{et al}~\cite{11},
in which $\rho_{\mathrm{o}}$ (induced by impurity) and
$\rho_{\mathrm{xxT}}$ (contributed by phonon) are considered to
have different contributions in the skew scattering term. For
L1(0) FePt films, one has
$\rho_{\mathrm{xyo}}=a_{\mathrm{o}}'\rho_{\mathrm{o}}+a_{\mathrm{o}}''\rho_{\mathrm{xxT}}+
b_{\mathrm{o}}'\rho_{\mathrm{xx}}^{\mathrm{2}}$. As shown in
Fig.~\ref{Fig3}(b), the values of $b_{\mathrm{o}}$ and
$b_{\mathrm{o}}'$ are close to each other for high $S$. This is
possibly because when $\rho_{\mathrm{o}}$ becomes small, the
difference between the new and the conventional approaches becomes
negligible as observed in Fe films~\cite{11}. More importantly,
$b_{\mathrm{o}}'$ also exhibits non-monotonic variation with $S$
similarly to $b_{\mathrm{o}}$. Finally, the AHE of L1(0) FePt
films with $S=0.8$ has very recently been studied~\cite{13}, in
which values of $a$ and $b$ are dramatically different from the
present results of $a_{\mathrm{o}}$ and $b_{\mathrm{o}}$, which is
likely because all data in Ref.[16] were analyzed from
$\rho_{\mathrm{xy}}$ instead of $f(T)$ independent $\rho_{\mathrm{xyo}}$.\\
\indent It is easy to understand the non-monotonic variations of
$a_{\mathrm{o}}$ and $b_{\mathrm{o}}$ according to theoretical
models about the AHC~\cite{6,14,15} under the assumption that
$\Delta_{\mathrm{SOC}}$ increases monotonically with $S$, i.e.,
the SOC constant $\xi$ of Pt and Fe atoms is\emph{ equivalently}
enhanced during the phase transformation. Here, the effect of the
exchange split energy on the AHC can be neglected although the AHC
arises from the interplay between the SOC and the exchange split.
This is because the effective spin magnetic moment does not change
much during phase transformation and thus the exchange splitting
energy between spin-up and spin-down bands is expected to change
little with $S$~\cite{39,40}. Firstly, Sinitsyn \emph{et al} have
studied the AHC dependence on $\xi$ in the two-dimensional Dirac
model system by a modified semiclassical transport
approach~\cite{14}. According to Eq.72 in this literature,
$\sigma_{\mathrm{sko}}$ has a minimum at
$\Delta_{\mathrm{gap}}/vk_{\mathrm{F}}\simeq0.5$ with
$\Delta_{\mathrm{gap}}\propto\xi^{\mathrm{2}}$ and $v$ being the
model parameter~\cite{13,38}. Alternatively, in the framework of
quantum transport theory, W$\ddot{\mathrm{o}}$lfle and Muttalib
have studied the AHE in (quasi) two-dimensional disordered
metallic band ferromagnet~\cite{15}. According to Eq.12 in
Ref.[9], $\sigma_{\mathrm{sko}}$ achieves a minimum at
$\xi/\xi_{\mathrm{o}}\simeq0.8$ with $\xi_{\mathrm{o}}$ being the
bare one. Since $\sigma_{\mathrm{xx}}$ is independent of $\xi$,
the non-monotonic variation of $a_{\mathrm{o}}$ in
Fig.~\ref{Fig3}(a) is easily understood.\\
\indent Secondly, as the sum of $\sigma_{\mathrm{sjo}}$ and
$\sigma_{\mathrm{into}}$, the non-monotonic variation of
$b_{\mathrm{o}}$ in Fig.~\ref{Fig3}(b) further confirms above
assumption. On one hand, $\sigma_{\mathrm{into}}$, which exists in
perfect crystals, is theoretically predicted to increase linearly
at small $\xi$ and to reach saturation for large $\xi$ as shown by
Fig.4 in Ref.[7] and Eq.58 in Ref.[8]. It is indirectly confirmed
by the Kerr rotation and the ellipticity enhancement in L1(0) FePt
films, compared with the disordered FePt films, as shown in
Fig.~\ref{Fig4}~\cite{16}. By utilizing sum rules for the optical
constants~\cite{26}, the intrinsic AHC (at the circle frequency
$\omega=0$) is suggested to be enhanced because the intrinsic AHC
has the same origin (both spin polarization and SOC) as the
MOKE~\cite{6}. On the other hand, since $\sigma_{\mathrm{sjo}}$
obeys the $\xi$ dependence similar to that of $a_{\mathrm{o}}$ as
shown by Eq.76 in Ref.[8] and Eq.17 in Ref.[9], it also changes
non-monotonically with $S$. Furthermore, it should be pointed out
that the impurity state can be excluded for the non-monotonic
variation trends of $a_{\mathrm{o}}$ and $b_{\mathrm{o}}$. Since
$\sigma_{\mathrm{sko}}$ and $\sigma_{\mathrm{xx}}$ are both
inversely proportional to the impurity concentration, as the ratio
$\sigma_{\mathrm{sko}}/\sigma_{\mathrm{xx}}$, $a_{\mathrm{o}}$ is
independent of the impurity concentration. $\sigma_{\mathrm{sjo}}$
is also independent of the impurity concentration albeit it arises
from the interplay of the impurity scattering and the
SOC~\cite{12}. As a consequence of the SOC in ideal crystals,
$\sigma_{\mathrm{into}}$ is not related to the impurity state at
all. According to the theoretical prediction by Yao, Sinitsyn, and
W$\ddot{\mathrm{o}}$lfle \emph{et al}~\cite{6,14,15}, the present
experimental results indicate
that the $\Delta_{\mathrm{SOC}}$ increases with increasing $S$.\\
\indent Theoretical calculations are encouraged to address
observed features of the AHC in L1(0) FePt films with varying $S$.
Here, well defined crystalline structure in L1(0) FePt films
favors direct comparison between experiments and calculations.
Although $\sigma_{\mathrm{int}}$ is proportional to
$M_{\mathrm{S}}$ in the temperature regime from 0 K to 300 K, the
SOC effect on the AHC cannot be taken into account by perturbation
approach because $a_{\mathrm{o}}$ and $b_{\mathrm{o}}$ vary
non-monotonically with $\xi$ as discussed above. Since
$M_{\mathrm{S}}$ is shown to obey linear dependence on
$T^{\mathrm{2}}$, the $M_{\mathrm{S}}$ reduction is proposed to be
caused by long-wavelength, low frequency fluctuation of spin
orientation at finite temperatures by Zeng \emph{et al}~\cite{9}.
Accordingly, this difficulty in theory is overcomed. For L1(0)
FePt films with small $S$, $\sigma_{\mathrm{sj}}$ also shows the
linear dependence on $M_{\mathrm{S}}$, however, different from
observations of Mn$_{\mathrm{5}}$Ge$_{\mathrm{3}}$ where
$\sigma_{\mathrm{sj}}$ is neglected. Furthermore, the increase of
$\Delta_{\mathrm{SOC}}$ with $S$ indicates the enhancement of
orbital polarization of Pt atoms due to chemical long range
ordering because the SOC at Pt sites plays a dominant role in the
magnetism of L1(0) FePt alloys~\cite{20,27}, although the orbital
polarization of Fe atoms is observed to be enhanced by about 300
$\%$~\cite{39,40}. In calculations, the lattice distortion and the
chemical ordering should be taken into account because the former
and the latter ones have great impact on the orbital polarization
in the magnitude and the anisotropic distribution, respectively.\\
 \indent In conclusion, it is likely the first time to have studied the AHC
dependence on $\Delta_{\mathrm{SOC}}$ in experiments, by employing
L1(0) FePt films. As the ratio $\rho_{\mathrm{xy}}/f(T)$,
$\rho_{\mathrm{xyo}}$ can be parameterized by $\rho_{\mathrm{xyo}}
=a_{\mathrm{o}}\rho_{\mathrm{xx}}+b_{\mathrm{o}}\rho_{\mathrm{xx}}^{\mathrm{2}}$.
Accordingly, $\rho_{\mathrm{xy}}$, $\sigma_{\mathrm{xy}}$,
$\sigma_{\mathrm {sk}}$, $\sigma_{\mathrm{int}}$, and in
particular $\sigma_{\mathrm{sj}}$ are all proportional to
$M_{\mathrm{S}}$ as a function of temperature in the regime of
0-300 K. It is interesting to find that $a_{\mathrm{o}}$ and
$b_{\mathrm{o}}$ change non-monotonically with $S$. Accordingly,
$\Delta_{\mathrm{SOC}}$ is verified to be enhanced in L1(0) FePt
films and to increase monotonically with $S$. The present state of
the art results will also be helpful to study electronic structure of L1(0) FePt films.\\
\indent Acknowledgements This work was supported by the National
Science Foundation of China Grant Nos. 50625102, 50871030, and
10974032, the National Basic Research Program of China under grant
No. 2009CB929201, 973-Project under Grant No. 2006CB921300,
and Shanghai Leading Academic Discipline Project under grant B113. \\

\newpage
\begin{figure}
\begin{center}
FIGURE CAPTIONS
\end{center}

\flushleft \indent Figure 1 Typical XRD spectra at high angles (a)
and XRR at low angles (b), Hall loops at 5 K(c), and temperature
dependence of the normalized spontaneous
magnetization, i.e., $f(T)$(d). In (d), the lines refer to linear fit results.\\

\indent Figure 2 $\rho_{\mathrm{xx}}$(a) and
$\rho_{\mathrm{xy}}$(b) versus temperature for FePt films with
various $S$, the ratio ($\alpha$) of
$\rho_{\mathrm{xy}}/\rho_{\mathrm{xx}}$(black squares) and
$\rho_{\mathrm{xyo}}/\rho_{\mathrm{xx}}$ (red circles) versus
$\rho_{\mathrm{xx}}$ for $S=0.86$ (c) and 0 (d). Here, symbols in
(a)-(d) refer to measured results, solid lines for
$\rho_{\mathrm{xyo}}/\rho_{\mathrm{xx}}$ curves in (c) and (d) to
fitted results by the linear function, and other lines in
(a)-(d) serve a guide to the eye.\\

\indent Figure 3 $a_{\mathrm{o}}$(a), $b_{\mathrm{o}}$ and
$b_{\mathrm{o}}'$ (b) as a function of $S$, which are fitted from
the curves of $\rho_{\mathrm{xyo}}/\rho_{\mathrm{xx}}$ versus
$\rho_{\mathrm{ xx}}$.\\

\indent Figure 4 Polar Kerr rotation $\theta_{\mathrm{K}}$ (a) and
ellipticity $\epsilon_{\mathrm{K}}$ (b) spectra of FePt films with
$S=0$ and 0.86. During measurements, the samples were in the
saturation
state. Lines serve a guide to the eye.\\

\end{figure}

\newpage
\begin{figure}[p]
\begin{center}
\resizebox*{6 in}{!}{\includegraphics*{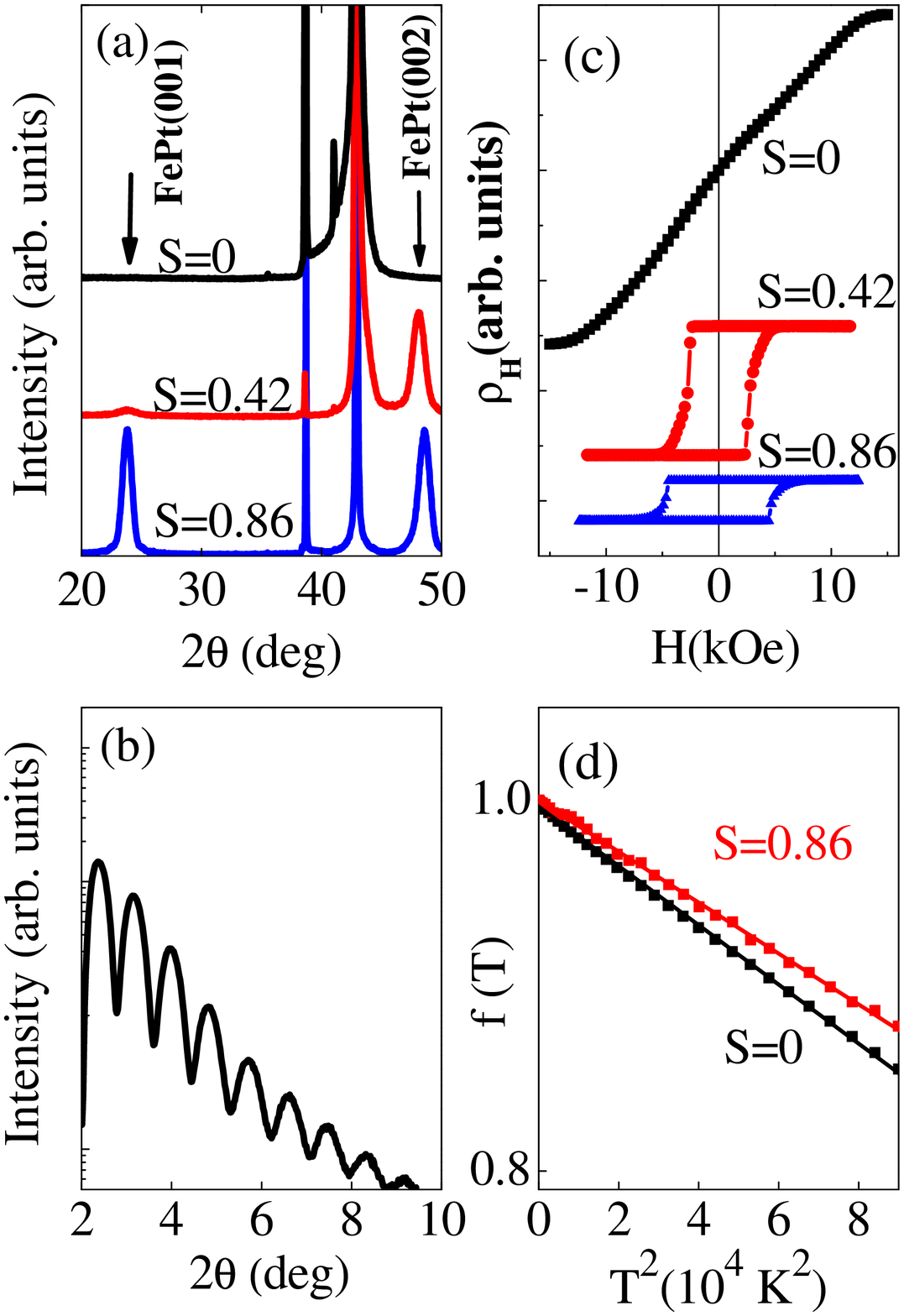}} \caption{}
\label{Fig1}
\end{center}
\end{figure}

\begin{figure}[p]
\begin{center}
\resizebox*{6 in}{!}{\includegraphics*{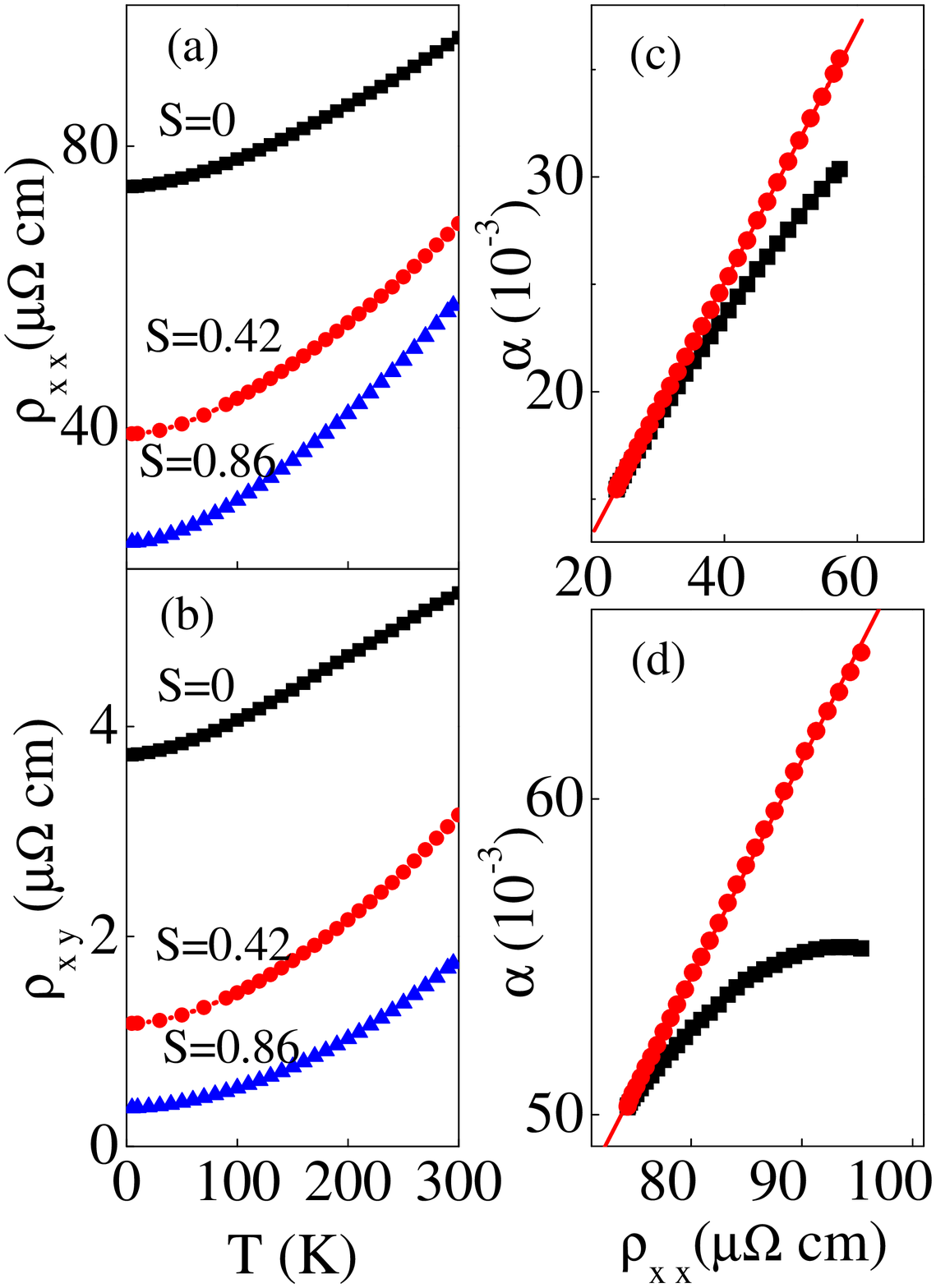}}
\caption{}\label{Fig2}
\end{center}
\end{figure}

\begin{figure}[p]
\begin{center}
\resizebox*{6 in}{!}{\includegraphics*{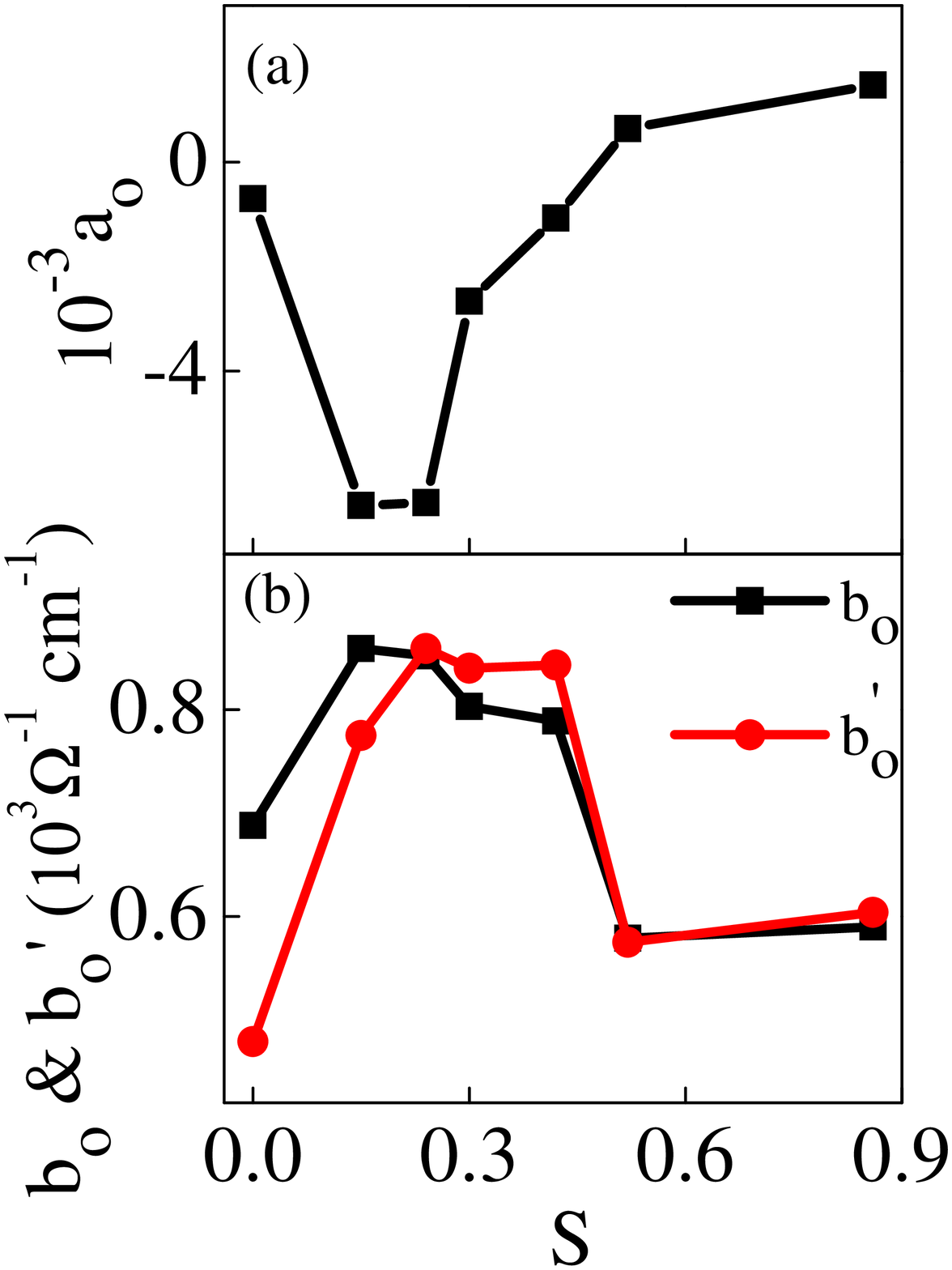}}
\caption{}\label{Fig3}
\end{center}
\end{figure}

\begin{figure}[p]
\begin{center}
\resizebox*{6 in}{!}{\includegraphics*{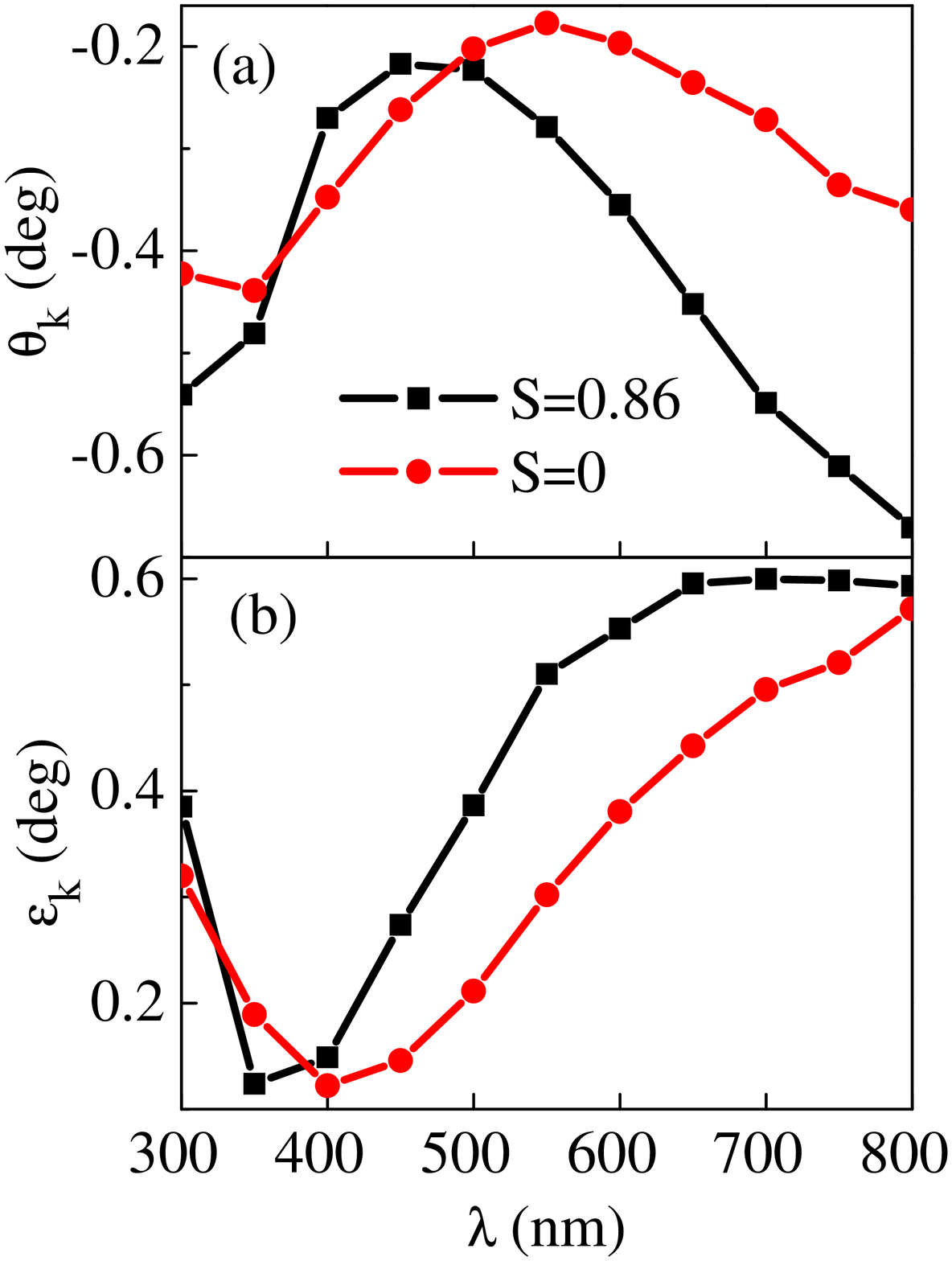}}
\caption{}\label{Fig4}
\end{center}
\end{figure}

\end{document}